\documentclass[11pt,onecolumn,draftcls]{IEEEtran}
\newtheorem{theorem}{Theorem}
\newtheorem{proposition}{Proposition}
\newtheorem{corollary}{Corollary}
\newtheorem{definition}{Definition}
\newtheorem{lemma}{Lemma}

\newcommand{\df}{\stackrel{\mbox{\scriptsize def}}{=}}

\newcommand{\rk}{\mathrm{rk}}

\newcommand{\dr}{d_{\mbox{\tiny{R}}}}

\newcommand{\vspan}[1]{\left< #1 \right>}

\usepackage{amsmath,amssymb,cite,subfigure}
\usepackage[dvips]{graphicx}

\begin{document}
\title{MacWilliams Identity for Codes with the Rank Metric}
\author{\authorblockN{Maximilien Gadouleau and Zhiyuan Yan}\\
\authorblockA{Department of Electrical and Computer Engineering\\
Lehigh University, Bethlehem, Pennsylvania 18015, USA\\
E-mail: \{magc, yan\}@lehigh.edu}
\thanks{The material in this paper was presented in part at the IEEE International Symposium on Information Theory, Nice, France,
June 24--29, 2007.}
}
\maketitle

\begin{abstract}
The MacWilliams identity, which relates the weight distribution of a
code to the weight distribution of its dual code, is useful in
determining the weight distribution of codes. In this paper, we
derive the MacWilliams identity for linear codes with the rank
metric, and our identity has a different form than that by Delsarte.
Using our MacWilliams identity, we also derive related identities
for rank metric codes. These identities parallel the binomial and
power moment identities derived for codes with the Hamming metric.
\end{abstract}

\section{Introduction}\label{sec:introduction}
The MacWilliams identity for codes with the Hamming metric
\cite{macwilliams_77}, which relates the Hamming weight distribution
of a code to the weight distribution of its dual code, is useful in
determining the Hamming weight distribution of codes. This is
because if the dual code has a small number of codewords or
equivalence classes of codewords under some know permutation group,
its weight distribution can be obtained by exhaustive examination.
It also leads to other identities for the weight distribution such
as the Pless identities \cite{macwilliams_77, pless_63}.

Although the rank has long been known to be a metric implicitly and
explicitly (see, for example, \cite{Hua51}), the rank metric was
first considered for error control codes (ECCs) by Delsarte
\cite{delsarte_78}. The potential applications of rank metric codes
to wireless communications \cite{tarokh_98,lusina_it03}, public-key
cryptosystems \cite{gabidulin_lncs91}, and storage equipments
\cite{gabidulin_pit0285,roth_it91} have motivated a steady stream of
works \cite{gabidulin_pit0185, gabidulin_pit0285, roth_it91,
chen_mn96, roth_it97, vasantha_gs99, berger_it03, gabidulin_isit05,
kshevetskiy_isit05, gadouleau_globecom06, gadouleau_itw06,
gadouleau_it06, loidreau_07} that focus on their properties. The
majority of previous works focus on rank distance properties, code
construction, and efficient decoding of rank metric codes, and the
seminal works in \cite{delsarte_78, gabidulin_pit0185, roth_it91}
have made significant contribution to these topics. Independently in
\cite{delsarte_78, gabidulin_pit0185, roth_it91}, a Singleton bound
(up to some variations) on the minimum rank distance of codes was
established, and a class of codes achieving the bound with equality
was constructed. We refer to this class of codes as Gabidulin codes
henceforth. In \cite{delsarte_78, gabidulin_pit0185}, analytical
expressions to compute the weight distribution of linear codes
achieving the Singleton bound with equality were also derived. In
\cite{gabidulin_pit0285}, it was shown that Gabidulin codes are
optimal for correcting crisscross errors (referred to as
lattice-pattern errors in \cite{gabidulin_pit0285}). In
\cite{roth_it91}, it was shown that Gabidulin codes are also optimal
in the sense of a Singleton bound in crisscross weight, a metric
considered in \cite{roth_it91,roth_it97,SE05} for crisscross errors.
Decoding algorithms were introduced for Gabidulin codes in
\cite{gabidulin_pit0185, roth_it91, richter_isit04, loidreau_05}.

In \cite{delsarte_78}, the counterpart of the MacWilliams identity,
which relates the rank distance enumerator of a code to that of its
dual code, was established using association schemes. However,
Delsarte's work lacks an expression of the rank weight enumerator of
the dual code as a functional transformation of the enumerator of
the code. In \cite{grant_allerton05, grant_07}, Grant and Varanasi
defined a {\em different} rank weight enumerator and established a
functional transformation between the rank weight enumerator of a
code and that of its dual code.

In this paper we show that, similar to the MacWilliams identity for
the Hamming metric, the rank weight distribution of any linear code
can be expressed as a functional transformation of that of its dual
code. It is remarkable that our MacWilliams identity for the rank
metric has a similar form to that for the Hamming metric. Similarly,
an intermediate result of our proof is that the rank weight
enumerator of the dual of any vector depends on only the rank weight
of the vector and is related to the rank weight enumerator of a
maximum rank distance (MRD) code. We also derive additional
identities that relate moments of the rank weight distribution of a
linear code to those of its dual code.

Our work in this paper differs from those in \cite{delsarte_78,
grant_allerton05, grant_07} in several aspects:
\begin{itemize}
    \item In this paper, we consider a rank weight enumerator
    different from that in \cite{grant_allerton05, grant_07}, and
    solve the original problem of determining the functional
    transformation of rank weight enumerators between dual codes
    as defined by Delsarte.

    \item Our proof, based
    on character theory, does not require the use of association schemes
    as in \cite{delsarte_78} or combinatorial arguments as in
    \cite{grant_allerton05, grant_07}.

    \item In \cite{delsarte_78}, the MacWilliams identity is given between the
    rank distance enumerator sequences of two dual array codes using the
    generalized Krawtchouk polynomials. Our identity is equivalent to
    that in \cite{delsarte_78} for linear rank metric codes, although
    our identity is expressed using different parameters which are shown
    to be the generalized Krawtchouk polynomials as well. We also
    present this identity in the form of a functional transformation
    (cf. Theorem~\ref{th:MacWilliams}). In such a form, the MacWilliams
    identities for both the rank and the Hamming metrics are similar to
    each other.

    \item The functional transformation form allows us to
    derive further identities (cf.
    Section~\ref{sec:moments}) between the rank weight distribution of linear dual
    codes. We would like to stress that the identities between the
    moments of the rank distribution proved in this paper are novel and
    were not considered in the aforementioned papers.
\end{itemize}

We remark that both the matrix form \cite{delsarte_78,roth_it91} and
the vector form \cite{gabidulin_pit0185} for rank metric codes have
been considered in the literature. Following
\cite{gabidulin_pit0185}, in this paper the vector form over
$\mathrm{GF}(q^m)$ is used for rank metric codes although their rank
weight is defined by their corresponding codematrices over
$\mathrm{GF}(q)$ \cite{gabidulin_pit0185}. The vector form is chosen
in this paper since our results and their derivations for rank
metric codes can be readily related to their counterparts for
Hamming metric codes.

The rest of the paper is organized as follows.
Section~\ref{sec:preliminaries} reviews some necessary background.
In Section~\ref{sec:rank}, we establish the MacWilliams identity for
the rank metric. We finally study the moments of the rank
distributions of linear codes in Section~\ref{sec:moments}.

\section{Preliminaries}\label{sec:preliminaries}
\subsection{Rank metric, MRD codes, and rank weight enumerator}\label{sec:rank_metric}
Consider an $n$-dimensional vector ${\bf x} = (x_0, x_1,\ldots,
x_{n-1}) \in \mathrm{GF}(q^m)^n$. The field $\mathrm{GF}(q^m)$ may
be viewed as an $m$-dimensional vector space over $\mathrm{GF}(q)$.
The rank weight of ${\bf x}$, denoted as $\rk({\bf x})$, is defined
to be the \emph{maximum} number of coordinates in ${\bf x}$ that are
linearly independent over $\mathrm{GF}(q)$ \cite{gabidulin_pit0185}.
Note that all ranks are with respect to $\mathrm{GF}(q)$ unless
otherwise specified in this paper. The coordinates of ${\bf x}$ thus
span a linear subspace of $\mathrm{GF}(q^m)$, denoted as
$\mathfrak{S}({\bf x})$, with dimension equal to $\rk({\bf x})$. For
all ${\bf x}, {\bf y}\in \mathrm{GF}(q^m)^n$, it is easily verified
that $\dr({\bf x},{\bf y})\df \rk({\bf x} - {\bf y})$ is a metric
over GF$(q^m)^n$ \cite{gabidulin_pit0185}, referred to as the
\emph{rank metric} henceforth. The {\em minimum rank distance} of a
code ${\mathcal C}$, denoted as $d_{\mbox{\tiny R}}({\mathcal C})$,
is simply the minimum rank distance over all possible pairs of
distinct codewords. When there is no ambiguity about ${\mathcal C}$,
we denote the minimum rank distance as $\dr$.

Combining the bounds in \cite{gabidulin_pit0185} and
\cite{loidreau_01} and generalizing slightly to account for
nonlinear codes, we can show that the cardinality $K$ of a code
${\mathcal C}$ over $\mathrm{GF}(q^m)$ with length $n$ and minimum
rank distance $\dr$ satisfies
\begin{equation}\label{eq:singleton3}
    K \leq \min \left\{q^{m(n-\dr+1)},q^{n(m-\dr+1)} \right\}.
\end{equation}
In this paper, we call the bound in (\ref{eq:singleton3}) the
Singleton bound for codes with the rank metric, and refer to codes
that attain the Singleton bound as maximum rank distance (MRD)
codes. We refer to MRD codes over $\mathrm{GF}(q^m)$ with length
$n\leq m$ and with length $n>m$ as Class-I and Class-II MRD codes
respectively. For any given parameter set $n$, $m$, and $\dr$,
explicit construction for linear or nonlinear MRD codes exists. For
$n\leq m$ and $\dr \leq n$, generalized Gabidulin codes
\cite{kshevetskiy_isit05} constitute a \emph{subclass} of linear
Class-I MRD codes. For $n>m$ and $\dr \leq m$, a Class-II MRD code
can be constructed by transposing a generalized Gabidulin code of
length $m$ and minimum rank distance $\dr$ over GF$(q^n)$, although
this code is not necessarily linear over GF$(q^m)$. When $n=lm$
($l\geq 2$), linear Class-II MRD codes of length $n$ and minimum
distance $\dr$ can be constructed by a cartesian product
$\mathcal{G}^l \df \mathcal{G} \times \ldots \times \mathcal{G}$ of
an $(m,k)$ linear Class-I MRD code $\mathcal{G}$ \cite{loidreau_01}.

For all ${\bf v} \in \mathrm{GF}(q^m)^n$ with rank weight $r$, the
rank weight function of ${\bf v}$ is defined as $f_{\mbox{\tiny
R}}({\bf v}) = y^{r}x^{n-r}$. Let ${\mathcal{C}}$ be a code of
length $n$ over $\mathrm{GF}(q^m)$. Suppose there are $A_i$
codewords in $\mathcal{C}$ with rank weight $i$ ($0 \leq i \leq n$),
then the rank weight enumerator of ${\mathcal{C}}$, denoted as
$W^{\mbox{\tiny R}}_{\mathcal{C}}(x,y)$, is defined to be
$$
W^{\mbox{\tiny R}}_{\mathcal{C}}(x,y) \df \sum_{{\bf v} \in
{\mathcal{C}}}f_{\mbox{\tiny R}}({\bf v}) = \sum_{i=0}^n A_i
y^{i}x^{n-i}.
$$

\subsection{Hadamard transform}\label{sec:hadamard}
\begin{definition}[\cite{macwilliams_77}]\label{def:chi}
Let $\mathbb{C}$ be the field of complex numbers. Let $a \in
\mathrm{GF}(q^m)$ and let $\{1,\alpha_1,\ldots,\alpha_{m-1} \}$ be a
basis set of $\mathrm{GF}(q^m)$. We thus have $a = a_0 + a_1
\alpha_1 + \ldots + a_{m-1}\alpha_{m-1}$, where $a_i \in
\mathrm{GF}(q)$ for $0 \leq i \leq m-1$. Finally, let $\zeta \in
\mathbb{C}$ be a primitive $q$-th root of unity, $\chi(a) \df
\zeta^{a_0}$ maps $\mathrm{GF}(q^m)$ to $\mathbb{C}$.
\end{definition}

\begin{definition}[Hadamard transform
\cite{macwilliams_77}]\label{def:hadamard} For a mapping $f$ from
$\mathrm{GF}(q^m)^n$ to $\mathbb{C}$, the {\em Hadamard transform}
of $f$, denoted as $\hat{f}$, is defined to be
\begin{equation}\label{eq:hadamard}
    \hat{f}({\bf v}) \df \sum_{{\bf u} \in \mathrm{GF}(q^m)^n} \chi({\bf u} \cdot {\bf
    v}) f({\bf u}),
\end{equation} where ${\bf u} \cdot {\bf
    v}$ denotes the inner product of ${\bf u}$ and ${\bf v}$.
\end{definition}

\subsection{Notations}\label{sec:notations}
In order to simplify notations, we shall occasionally denote the
vector space $\mathrm{GF}(q^m)^n$ as $F$. We denote the number of
vectors of rank $u$ ($0 \leq u \leq \min\{m,n\}$) in
$\mathrm{GF}(q^m)^n$ as $N_u(q^m,n)$. It can be shown that
$N_u(q^m,n) = {n \brack u} \alpha(m,u)$ \cite{gabidulin_pit0185},
where $\alpha(m,0) \df 1$ and $\alpha(m,u) \df
\prod_{i=0}^{u-1}(q^m-q^i)$ for $u \geq 1$. The ${n \brack u}$ term
is often referred to as a Gaussian polynomial~\cite{andrews},
defined as ${n \brack u} \df \alpha(n,u)/\alpha(u,u)$. Note that ${n
\brack u}$ is the number of $u$-dimensional linear subspaces of
$\mathrm{GF}(q)^n$. We also define $\beta(m,0) \df 1$ and
$\beta(m,u) \df \prod_{i=0}^{u-1} {m-i \brack 1}$ for $u \geq 1$.
These terms are closely related to Gaussian polynomials: $\beta(m,u)
= {m \brack u} \beta(u,u)$ and $\beta(m+u,m+u) = {m+u \brack u}
\beta(m,m) \beta(u,u)$. Finally, $\sigma_i \df \frac{i(i-1)}{2}$ for
$i \geq 0$.

\section{MacWilliams identity for the rank metric}\label{sec:rank}
\subsection{$q$-product, $q$-transform, and $q$-derivative}
\label{sec:star_prod_q-derivative} In order to express the
MacWilliams identity in polynomial form as well as to derive other
identities, we introduce several operations on homogeneous
polynomials.

Let $a(x,y;m) = \sum_{i=0}^r a_{i}(m) y^i x^{r-i}$ and $b(x,y;m) =
\sum_{j=0}^s b_{j}(m) y^j x^{s-j}$ be two homogeneous polynomials in
$x$ and $y$ of degrees $r$ and $s$ respectively with coefficients
$a_i(m)$ and $b_j(m)$ respectively. $a_i(m)$ and $b_j(m)$ for $i,j
\geq 0$ in turn are real functions of $m$, and are assumed to be
zero unless otherwise specified.

\begin{definition}[$q$-product]\label{def:star_prod}
The {\em $q$-product} of $a(x,y;m)$ and $b(x,y;m)$ is defined to be
the homogeneous polynomial of degree $(r+s)$ $c(x,y;m) \df a(x,y;m)
* b(x,y;m) = \sum_{u=0}^{r+s} c_{u}(m) y^u x^{r+s-u}$, with
\begin{equation}
    c_{u}(m) = \sum_{i=0}^u q^{is} a_{i}(m) b_{u-i}(m-i).
\label{eq:q-product}\end{equation}

We shall denote the $q$-product by $*$ henceforth. For $n \geq 0$
the $n$-th $q$-power of $a(x,y;m)$ is defined recursively:
$a(x,y;m)^{[0]} = 1$ and $a(x,y;m)^{[n]} =
a(x,y;m)^{[n-1]}*a(x,y;m)$ for $n \geq 1$.
\end{definition}

We provide some examples to illustrate the concept. It is easy to
verify that $x * y = yx$, $y * x = qyx$, $yx * x = q yx^2$, and $yx
* (q^m-1)y = (q^m-q)y^2x$. Note that $x * y \ne y * x$. It is easy
to verify that the $q$-product is neither commutative nor
distributive in general. However, it is commutative and distributive
in some special cases as described below.

\begin{lemma}\label{lemma:properties_star_prod}
Suppose $a(x,y;m) = a$ is a constant independent from $m$, then
$a(x,y;m)*b(x,y;m) = b(x,y;m)*a(x,y;m) = ab(x,y;m)$. Also, if
$\deg[c(x,y;m)] = \deg[a(x,y;m)]$, then
$[a(x,y;m)+c(x,y;m)]*b(x,y;m) = a(x,y;m)*b(x,y;m) +
c(x,y;m)*b(x,y;m)$, and $b(x,y;m) * [a(x,y;m)+c(x,y;m)] =
b(x,y;m)*a(x,y;m) + b(x,y;m)*c(x,y;m)$.
\end{lemma}

The homogeneous polynomials $a_l(x,y;m) \df [x+(q^m-1)y]^{[l]}$ and
$b_l(x,y;m) \df (x-y)^{[l]}$ are very important to our derivations
below. The following lemma provides the analytical expressions of
$a_l(x,y;m)$ and $b_l(x,y;m)$.

\begin{lemma}\label{lemma:special_prod}
For $l\geq 0$, we have $y^{[l]} = q^{\sigma_l}y^l$ and $x^{[l]} =
x^l$. Furthermore,
\begin{eqnarray}
    \label{eq:x+y^s}
    a_l(x,y;m) = \sum_{u=0}^l {l \brack u} \alpha(m,u) y^u
    x^{l-u},\\
    \label{eq:x-y^r}
    b_l(x,y;m) = \sum_{u=0}^l {l \brack u} (-1)^u q^{\sigma_u} y^u x^{l-u}.
\end{eqnarray}
\end{lemma}

Note that $a_l(x,y;m)$ is the rank weight enumerator of
$\mathrm{GF}(q^m)^l$. The proof of Lemma~\ref{lemma:special_prod},
which goes by induction on $l$, is easy and hence omitted.

\begin{definition}[$q$-transform]\label{def:star_transform}
We define the {\em $q$-transform} of $a(x,y;m)= \sum_{i=0}^r
a_{i}(m) y^i x^{r-i}$ as the homogeneous polynomial $\bar{a}(x,y;m)=
\sum_{i=0}^r a_{i}(m) y^{[i]}* x^{[r-i]}$.
\end{definition}

\begin{definition}[$q$-derivative \cite{gasper_book04}]
\label{def:q-derivative} For $q \geq 2$, the $q$-derivative at $x
\neq 0$ of a real-valued function $f(x)$ is defined as
$$
    f^{(1)}(x) \df \frac{f(qx)-f(x)}{(q-1)x}.
$$
\end{definition}

For any real number $a$, $[f(x)+ag(x)]^{(1)} = f^{(1)}(x) +
ag^{(1)}(x)$ for $x \neq 0$. For $\nu \geq 0$, we shall denote the
$\nu$-th $q$-derivative (with respect to $x$) of $f(x,y)$ as
$f^{(\nu)}(x,y)$. The $0$-th $q$-derivative of $f(x,y)$ is defined
to be $f(x,y)$ itself.

\begin{lemma}\label{lemma:special_q-d}
For $0 \leq \nu \leq l$, $(x^l)^{(\nu)} = \beta(l,\nu)x^{l-\nu}$.
The $\nu$-th $q$-derivative of $f(x,y)= \sum_{i=0}^r f_i y^i
x^{r-i}$ is given by $f^{(\nu)}(x,y) = \sum_{i=0}^{r-\nu} f_i
\beta(i,\nu) y^i x^{r-i-\nu}$. Also,
\begin{eqnarray}
    \label{eq:al_nu}
    a_l^{(\nu)}(x,y;m) &=& \beta(l,\nu) a_{l-\nu}(x,y;m)\\
    \label{eq:bl_nu}
    b_l^{(\nu)}(x,y;m) &=& \beta(l,\nu) b_{l-\nu}(x,y;m).
\end{eqnarray}
\end{lemma}

The proof of Lemma~\ref{lemma:special_q-d}, which goes by induction
on $\nu$, is easy and hence omitted.

\begin{lemma}[Leibniz rule for the $q$-derivative]\label{lemma:Leibniz_x}
For two homogeneous polynomials $f(x,y)$ and $g(x,y)$ with degrees
$r$ and $s$ respectively, the $\nu$-th ($\nu \geq 0$) $q$-derivative
of their $q$-product is given by
\begin{equation}
    \label{eq:leibniz_nu}
    \left[f(x,y)*g(x,y)\right]^{(\nu)} = \sum_{l=0}^{\nu} {\nu \brack l}
    q^{(\nu-l)(r-l)} f^{(l)}(x,y)*g^{(\nu-l)}(x,y).
\end{equation}
\end{lemma}

The proof of Lemma~\ref{lemma:Leibniz_x} is given in
Appendix~\ref{app:lemma:leibniz_x}.

The $q^{-1}$-derivative is similar to the $q$-derivative.
\begin{definition}[$q^{-1}$-derivative]\label{def:q-1_derivative}
For $q \geq 2$, the $q^{-1}$-derivative at $y \neq 0$ of a
real-valued function $g(y)$ is defined as
$$
    g^{\{1\}}(y) \df \frac{g(q^{-1}y) - g(y)}{(q^{-1} - 1)y}.
$$
\end{definition}

For any real number $a$, $[f(y)+ag(y)]^{\{1\}} = f^{\{1\}}(y) +
ag^{\{1\}}(y)$ for $y \neq 0$. For $\nu \geq 0$, we shall denote the
$\nu$-th $q^{-1}$-derivative (with respect to $y$) of $g(x,y)$ as
$g^{\{\nu\}}(x,y)$. The $0$-th $q^{-1}$-derivative of $g(x,y)$ is
defined to be $g(x,y)$ itself.

\begin{lemma}\label{lemma:special_q-1_d}
For $0 \leq \nu \leq l$, the $\nu$-th $q^{-1}$-derivative of $y^l$
is $(y^l)^{\{ \nu \}} = q^{\nu(1-n) + \sigma_\nu} \beta(l,\nu)
y^{l-\nu}$. Also,
\begin{eqnarray}
    \label{eq:al_nu_q-1}
    a_l^{\{ \nu \}}(x,y;m) &=& \beta(l,\nu) q^{-\sigma_\nu} \alpha(m,\nu)
    a_{l-\nu}(x,y;m-\nu)\\
    \label{eq:bl_nu_q-1}
    b_l^{\{ \nu \}}(x,y;m) &=& (-1)^{\nu}\beta(l,\nu)b_{l-\nu}(x,y;m).
\end{eqnarray}
\end{lemma}

The proof of Lemma~\ref{lemma:special_q-1_d} is similar to that of
Lemma~\ref{lemma:special_q-d} and is hence omitted.

\begin{lemma}[Leibniz rule for the $q^{-1}$-derivative]\label{lemma:Leibniz_y}
For two homogeneous polynomials $f(x,y;m)$ and $g(x,y;m)$ with
degrees $r$ and $s$ respectively, the $\nu$-th ($\nu \geq 0$)
$q^{-1}$-derivative of their $q$-product is given by
\begin{equation}\label{eq:leibniz_y_nu}
    [f(x,y;m)*g(x,y;m)]^{\{ \nu \}} = \sum_{l=0}^{\nu} {\nu \brack l}
    q^{l(s-\nu+l)} f^{\{ l \}}(x,y;m)*g^{\{ \nu-l \}}(x,y;m-l).
\end{equation}
\end{lemma}

The proof of Lemma~\ref{lemma:Leibniz_y} is given in
Appendix~\ref{app:lemma:leibniz_y}.

\subsection{The dual of a vector}\label{sec:dual_v_rank}
As an important step toward our main result, we derive the rank
weight enumerator of $\vspan{{\bf v}}^\perp$, where ${\bf v} \in
\mathrm{GF}(q^m)^{n}$ is an arbitrary vector and $\vspan{{\bf v}}\df
\left\{a{\bf v}: a \in \mathrm{GF}(q^m)\right\}$. Note that
$\vspan{{\bf v}}$ can be viewed as an $(n,1)$ linear code over
$\mathrm{GF}(q^m)$ with a generator matrix ${\bf v}$. It is
remarkable that the rank weight enumerator of $\vspan{{\bf
v}}^\perp$ depends on only the rank of ${\bf v}$.

Berger \cite{berger_it03} has determined that linear isometries for
the rank distance are given by the scalar multiplication by a
non-zero element of $\mathrm{GF}(q^m)$, and multiplication on the
right by an nonsingular matrix ${\bf B} \in \mathrm{GF}(q)^{n \times
n}$. We say that two codes $C$ and $C'$ are rank-equivalent if there
exists a linear isometry $f$ for the rank distance such that $f(C) =
C'$.

\begin{lemma}\label{lemma:rk_dual_vector}
Suppose ${\bf v}$ has rank $r \geq 1$, Then $\mathcal{L} =
\vspan{{\bf v}}^\perp$ is rank-equivalent to $\mathcal{C} \times
\mathrm{GF}(q^m)^{n-r}$, where $\mathcal{C}$ is an $(r,r-1,2)$ MRD
code and $\times$ denotes cartesian product.
\end{lemma}

\begin{proof}
We can express ${\bf v}$ as ${\bf v} = \bar{\bf v} {\bf B}$, where
$\bar{\bf v} = (v_0,\ldots,v_{r-1},0\ldots,0)$ has rank $r$, and
${\bf B} \in \mathrm{GF}(q)^{n \times n}$ has full rank. Remark that
$\bar{\bf v}$ is the parity-check of $\mathcal{C} \times
\mathrm{GF}(q^m)^{n-r}$, where $\mathcal{C} =
\vspan{(v_0,\ldots,v_{r-1})}^\perp$ is an $(r,r-1,2)$ MRD code. It
can be easily checked that ${\bf u} \in \mathcal{L}$ if and only if
$\bar{\bf u} \df {\bf u}{\bf B}^T \in \vspan{\bar{\bf v}}^\perp$.
Therefore, $\vspan{\bar{\bf v}}^\perp = \mathcal{L} {\bf B}^T$, and
hence $\mathcal{L}$ is rank-equivalent to $\vspan{\bar{\bf v}}^\perp
= \mathcal{C} \times \mathrm{GF}(q^m)^{n-r}$.
\end{proof}

We hence derive the rank weight enumerator of an $(r,r-1,2)$ MRD
code. Note that the rank weight distribution of linear Class-I MRD
codes has been derived in \cite{delsarte_78,gabidulin_pit0185}.
However, we shall not use the result in
\cite{delsarte_78,gabidulin_pit0185}, and instead derive the rank
weight enumerator of an $(r,r-1,2)$ MRD code directly.

\begin{proposition}\label{prop:trivial_MRD}
Suppose ${\bf v}_r \in \mathrm{GF}(q^m)^r$ has rank $r$ ($0\leq r
\leq m$). The rank weight enumerator of $\mathcal{L}_r = \vspan{{\bf
v}}^{\perp}$ depends on only $r$ and is given by
\begin{equation}
    W_{\mathcal{L}_r}^{\mbox{\tiny R}}(x,y) =
     q^{-m}\left\{ \left[x+(q^m-1)y \right]^{[r]} + (q^m-1)(x-y)^{[r]}\right\}.
\label{eq:trivial_MRD}\end{equation}
\end{proposition}

\begin{proof}
We first prove that the number of vectors with rank $r$ in
$\mathcal{L}_r$, denoted as $A_{r,r}$, depends only on $r$ and is
given by
\begin{equation}\label{eq:App}
    A_{r,r} = q^{-m}[\alpha(m,r) + (q^m-1)(-1)^r q^{\sigma_r}]
\end{equation}
by induction on $r$ ($r \geq 1$). Eq.~(\ref{eq:App}) clearly holds
for $r=1$. Suppose~(\ref{eq:App}) holds for $r = {\bar r}-1$.

We consider all the vectors ${\bf u} = (u_0,\ldots,u_{{\bar r}-1})
\in \mathcal{L}_{\bar r}$ such that the first ${\bar r}-1$
coordinates of ${\bf u}$ are linearly independent. Remark that
$u_{{\bar r}-1} = -v_{{\bar r}-1}^{-1} \sum_{i=0}^{{\bar r}-2}
u_iv_i$ is completely determined by $u_0,\ldots,u_{{\bar r}-2}$.
Thus there are $N_{{\bar r}-1}(q^m,{\bar r}-1) = \alpha(m,{\bar r}
-1)$ such vectors ${\bf u}$. Among these vectors, we will enumerate
the vectors ${\bf t}$ whose last coordinate is a linear combination
of the first ${\bar r}-1$ coordinates, i.e., ${\bf t} =
(t_0,\ldots,t_{{\bar r}-2},\sum_{i=0}^{{\bar r}-2} a_it_i) \in
\mathcal{L}_{\bar r}$ where $a_i \in \mathrm{GF}(q)$ for $0 \leq i
\leq {\bar r}-2$.

Remark that ${\bf t} \in \mathcal{L}_{\bar r}$ if and only if
$(t_0,\ldots,t_{{\bar r}-2}) \cdot (v_0 + a_0 v_{{\bar r}-1},
\ldots, v_{{\bar r}-2} + a_{{\bar r}-2} v_{{\bar r}-1}) = 0$. It is
easy to check that ${\bf v}({\bf a}) = (v_0 + a_0 v_{{\bar r}-1},
\ldots, v_{{\bar r}-2} + a_{{\bar r}-2} v_{{\bar r}-1})$ has rank
${\bar r}-1$. Therefore, if $a_0, \ldots, a_{{\bar r}-2}$ are fixed,
then there are $A_{{\bar r}-1,{\bar r}-1}$ such vectors ${\bf t}$.
Also, suppose $\sum_{i=0}^{{\bar r}-2} t_i v_i + v_{{\bar r}
-1}\sum_{i=0}^{{\bar r}-2} b_it_i = 0$. Hence $\sum_{i=0}^{{\bar r}
-2} (a_i -b_i)t_i=0$, which implies ${\bf a} = {\bf b}$ since
$t_i$'s are linearly independent. That is, $\vspan{{\bf v}({\bf
a})}^\perp \cap \vspan{{\bf v}({\bf b})}^\perp = \left\{{\bf
0}\right\}$ if ${\bf a} \neq {\bf b}$. We conclude that there are
$q^{{\bar r}-1} A_{{\bar r}-1,{\bar r}-1}$ vectors ${\bf t}$.
Therefore, $A_{{\bar r}, {\bar r}} = \alpha(m,{\bar r}-1) - q^{{\bar
r}-1} A_{{\bar r}-1,{\bar r}-1} = q^{-m} [\alpha(m,{\bar r}) +
(q^m-1)(-1)^{\bar r} q^{\sigma_{\bar r}}]$.

Denote the number of vectors with rank $p$ in $\mathcal{L}_r$ as
$A_{r,p}$. We have $A_{r,p} = {r \brack p}A_{p,p}$
\cite{gabidulin_pit0185}, and hence $A_{r,p} = {r \brack p} q^{-m}
[\alpha(m,p) + (q^m-1)(-1)^p q^{\sigma_p}]$. Thus, $W^{\mbox{\tiny
R}}_{\mathcal{L}_r}(x,y) = \sum_{p=0}^r A_{r,p} x^{r-p}y^p = q^{-m}
\Big\{ \left[x+(q^m-1)y \right]^{[r]} + (q^m-1)(x-y)^{[r]} \Big\}$.
\end{proof}

We comment that Proposition~\ref{prop:trivial_MRD} in fact provides
the rank weight distribution of any $(r,r-1,2)$ MRD code.

\begin{lemma}\label{lemma:rk_Asu1}
Let $\mathcal{C}_0 \subseteq \mathrm{GF}(q^m)^r$ be a linear code
with rank weight enumerator $W_{\mathcal{C}_0}^{\mbox{\tiny
R}}(x,y)$, and for $s \geq 0$, let $W_{\mathcal{C}_s}^{\mbox{\tiny
R}}(x,y)$ be the rank weight enumerator of ${\mathcal{C}_s} \df
\mathcal{C}_0 \times \mathrm{GF}(q^m)^s$. Then
$W_{\mathcal{C}_s}^{\mbox{\tiny R}}(x,y)$ is given by
\begin{equation}\label{eq:rk_Asu}
    W_{\mathcal{C}_s}^{\mbox{\tiny R}}(x,y) = W_{\mathcal{C}_0}^{\mbox{\tiny R}}(x,y) *
    \left[x+(q^m-1)y\right]^{[s]}.
\end{equation}
\end{lemma}

\begin{proof}
For $s \geq 0$, denote $W_{\mathcal{C}_s}^{\mbox{\tiny R}}(x,y) =
\sum_{u=0}^{r+s} B_{s,u} y^u x^{r+s-u}$. We will prove that
\begin{equation}\label{eq:Bsu}
    B_{s,u} = \sum_{i=0}^u q^{is} B_{0,i} {s \brack u-i}
    \alpha(m-i,u-i)
\end{equation}
by induction on $s$. Eq.~(\ref{eq:Bsu}) clearly holds for $s=0$. Now
assume (\ref{eq:Bsu}) holds for $s = {\bar s}-1$. For any ${\bf
x}_{\bar s} = (x_0,\ldots,x_{r+{\bar s}-1}) \in \mathcal{C}_{\bar
s}$, we define ${\bf x}_{{\bar s}-1} = (x_0,\ldots,x_{r+{\bar s}-2})
\in \mathcal{C}_{{\bar s}-1}$. Then $\rk({\bf x}_{\bar s}) = u$ if
and only if either $\rk({\bf x}_{{\bar s}-1}) = u$ and $x_{r+{\bar
s}-1} \in \mathfrak{S}({\bf x}_{{\bar s}-1})$ or $\rk({\bf x}_{{\bar
s}-1}) = u-1$ and $x_{r+{\bar s}-1} \notin \mathfrak{S} ({\bf
x}_{{\bar s}-1})$. This implies $B_{{\bar s},u} = q^u B_{{\bar
s}-1,u} + (q^m-q^{u-1})B_{{\bar s}-1,u-1}= \sum_{i=0}^u q^{i{\bar
s}} B_{0,i} {{\bar s} \brack u-i} \alpha(m-i,u-i)$.
\end{proof}

Combining Lemma~\ref{lemma:rk_dual_vector},
Proposition~\ref{prop:trivial_MRD}, and Lemma~\ref{lemma:rk_Asu1},
the rank weight enumerator of $\vspan{{\bf v}}^\perp$ can be
determined at last.

\begin{proposition}\label{prop:rk_W_L}
For ${\bf v} \in \mathrm{GF}(q^m)^n$ with rank $r \geq 0$, the rank
weight enumerator of $\mathcal{L} = \vspan{{\bf v}}^{\perp}$ depends
on only $r$, and is given by
\begin{equation}
    W_\mathcal{L}^{\mbox{\tiny R}}(x,y) = q^{-m} \left\{
    \left[x+(q^m-1)y\right]^{[n]} + (q^m-1) (x-y)^{[r]} *
    \left[x+(q^m-1)y\right]^{[n-r]} \right\}.
\end{equation}
\end{proposition}

\subsection{MacWilliams identity for the rank metric} \label{sec:theorem_rank}
Using the results in Section~\ref{sec:dual_v_rank}, we now derive
the MacWilliams identity for rank metric codes. Let $\mathcal{C}$ be
an $(n,k)$ linear code over $\mathrm{GF}(q^m)$, and let
$W_\mathcal{C}^{\mbox{\tiny R}}(x,y) = \sum_{i=0}^n A_i y^ix^{n-i}$
be its rank weight enumerator and
$W_{\mathcal{C}^{\perp}}^{\mbox{\tiny R}}(x,y) = \sum_{j=0}^n B_j
y^j x^{n-j}$ be the rank weight enumerator of its dual code
$\mathcal{C}^{\perp}$.

\begin{theorem}\label{th:MacWilliams}
For any $(n,k)$ linear code $\mathcal{C}$ and its dual code
$\mathcal{C}^{\perp}$ over $\mathrm{GF}(q^m)$,
\begin{equation}\label{eq:macwilliams}
    W_{\mathcal{C}^{\perp}}^{\mbox{\tiny R}}(x,y) = \frac{1}{|\mathcal{C}|}
    {\bar W}_\mathcal{C}^{\mbox{\tiny R}}\left(x+(q^m-1)y,x-y\right),
\end{equation}
where ${\bar W}_\mathcal{C}^{\mbox{\tiny R}}$ is the $q$-transform
of $W_\mathcal{C}^{\mbox{\tiny R}}$. Equivalently,
\begin{equation}\label{eq:macwilliams2}
    \sum_{j=0}^n B_j y^j x^{n-j} = q^{-mk}\sum_{i=0}^n A_i
    (x-y)^{[i]}* \left[x+(q^m-1)y\right]^{[n-i]}.
\end{equation}
\end{theorem}

\begin{proof}
We have $\rk(\lambda {\bf u}) = \rk({\bf u})$ for all $\lambda \in
\mathrm{GF}(q^m)^*$ and all ${\bf u} \in \mathrm{GF}(q^m)^n$. We
want to determine $\hat{f}_{\mbox{\tiny R}}({\bf v})$ for all ${\bf
v} \in \mathrm{GF}(q^m)^n$. By Definition~\ref{def:hadamard}, we can
split the summation in~(\ref{eq:hadamard}) into two parts:
$$
    \hat{f}_{\mbox{\tiny R}}({\bf v}) = \sum_{{\bf u} \in \mathcal{L}} \chi({\bf u} \cdot {\bf v})
    f_{\mbox{\tiny R}}({\bf u}) + \sum_{{\bf u} \in F \backslash \mathcal{L}} \chi({\bf u} \cdot
    {\bf v}) f_{\mbox{\tiny R}}({\bf u}),
$$
where $\mathcal{L} = \vspan{{\bf v}}^{\perp}$. If ${\bf u} \in
\mathcal{L}$, then $\chi({\bf u} \cdot {\bf v}) = 1$ by
Definition~\ref{def:chi}, and the first summation is equal to
$W^{\mbox{\tiny R}}_\mathcal{L}(x,y)$. For the second summation, we
divide vectors into groups of the form $\{\lambda {\bf u}_1\}$,
where $\lambda \in \mathrm{GF}(q^m)^*$ and ${\bf u}_1 \cdot {\bf v}
= 1$. We remark that for ${\bf u} \in F \backslash \mathcal{L}$ (see
\cite[Chapter 5, Lemma 9]{macwilliams_77})
$$
    \sum_{\lambda \in \mathrm{GF}(q^m)^*} \chi(\lambda{\bf u}_1 \cdot {\bf
    v})f_{\mbox{\tiny R}}(\lambda {\bf u}_1) =
    f_{\mbox{\tiny R}}({\bf u}_1)\sum_{\lambda \in \mathrm{GF}(q^m)^*} \chi(\lambda) =
    -f_{\mbox{\tiny R}}({\bf u}_1).
$$
Hence the second summation is equal to $-\frac{1}{q^m-1}
W^{\mbox{\tiny R}}_{F\backslash \mathcal{L}}(x,y)$. This leads to
$\hat{f}_{\mbox{\tiny R}}({\bf v}) = \frac{1}{q^m-1} [q^m
W^{\mbox{\tiny R}}_\mathcal{L}(x,y) - W^{\mbox{\tiny R}}_F(x,y)].$
Using $W_F^{\mbox{\tiny R}}(x,y) = [x+(q^m-1)y]^{[n]}$ and
Proposition~\ref{prop:rk_W_L}, we obtain
$\hat{f}_{\mbox{\tiny{R}}}({\bf v}) = (x-y)^{[r]}*
\left[x+(q^m-1)y\right]^{[n-r]}$, where $r=\rk({\bf v})$.

By \cite[Chapter 5, Lemma 11]{macwilliams_77}, any mapping $f$ from
$F$ to $\mathbb{C}$ satisfies $\sum_{{\bf v} \in
\mathcal{C}^{\perp}} f({\bf v}) = \frac{1}{|\mathcal{C}|} \sum_{{\bf
v} \in \mathcal{C}} \hat{f}({\bf v}).$ Applying this result to
$f_{\mbox{\tiny{R}}}({\bf v})$ and using
Definition~\ref{def:star_transform}, we obtain
(\ref{eq:macwilliams}) and (\ref{eq:macwilliams2}).
\end{proof}

Also, $B_j$'s can be explicitly expressed in terms of $A_i$'s.

\begin{corollary}\label{cor:krawtchouk}
We have
\begin{equation}\label{eq:B_A_krawtchouk}
    B_j = \frac{1}{|\mathcal{C}|} \sum_{i=0}^n A_i P_j(i;m,n),
\end{equation}
where
\begin{equation}\label{eq:krawtchouk2}
    P_j(i;m,n) \df \sum_{l=0}^j {i \brack l}{n-i \brack j-l}(-1)^l
    q^{\sigma_l}q^{l(n-i)}\alpha(m-l,j-l).
\end{equation}
\end{corollary}

\begin{proof}
We have $(x-y)^{[i]}*(x+(q^m-1)y)^{[n-i]} = \sum_{j=0}^n P_j(i;m,n)
y^jx^{n-j}$. The result follows Theorem~\ref{th:MacWilliams}.
\end{proof}

Note that although the analytical expression in
(\ref{eq:B_A_krawtchouk}) is similar to that in
\cite[(3.14)]{delsarte_78}, $P_j(i;m,n)$ in~(\ref{eq:krawtchouk2})
are different from $P_j(i)$ in \cite[(A10)]{delsarte_78} and their
alternative forms in \cite{delsarte_76}. We can show that

\begin{proposition}\label{prop:P=Q}
$P_j(x;m,n)$ in~(\ref{eq:krawtchouk2}) are the generalized
Krawtchouk polynomials.
\end{proposition}

The proof is given in Appendix~\ref{app:prop:P=Q}.
Proposition~\ref{prop:P=Q} shows that $P_j(x;m,n)$
in~(\ref{eq:krawtchouk2}) are an alternative form for $P_j(i)$ in
\cite[(A10)]{delsarte_78}, and hence our results in
Corollary~\ref{cor:krawtchouk} are equivalent to those in
\cite[Theorem~3.3]{delsarte_78}. Also, it was pointed out in
\cite{delsarte_76} that $\frac{P_j(x;m,n)}{P_j(0;m,n)}$ is actually
a basic hypergeometric function.

\section{Moments of the rank distribution}\label{sec:moments}
\subsection{Binomial moments of the rank distribution}\label{sec:moments_distrib}
In this section, we investigate the relationship between moments of
the rank distribution of a linear code and those of its dual code.
Our results parallel those in \cite[p. 131]{macwilliams_77}.

\begin{proposition}\label{prop:bm_x}
For $0 \leq \nu \leq n$,
\begin{equation}\label{eq:bm_x}
     \sum_{i=0}^{n-\nu} {n-i \brack \nu} A_i = q^{m(k-\nu)}
     \sum_{j=0}^{\nu} {n-j \brack n-\nu} B_j.
\end{equation}
\end{proposition}

\begin{proof}
First, applying Theorem~\ref{th:MacWilliams} to
$\mathcal{C}^{\perp}$, we obtain
\begin{equation}\label{eq:before_nu}
    \sum_{i=0}^n A_i y^i x^{n-i} = q^{m(k-n)} \sum_{j=0}^n B_j b_j(x,y;m)*a_{n-j}(x,y;m).
\end{equation}

Next, we apply the $q$-derivative with respect to $x$
to~(\ref{eq:before_nu}) $\nu$ times. By
Lemma~\ref{lemma:special_q-d} the left hand side (LHS) becomes
$\sum_{i=0}^{n-\nu} \beta(n-i,\nu) A_i y^i x^{n-i-\nu}$, while the
RHS reduces to $q^{m(k-n)}\sum_{j=0}^n B_j \psi_j(x,y)$ by
Lemma~\ref{lemma:Leibniz_x}, where
$$
    \psi_j(x,y) \df
    [b_j(x,y;m)*a_{n-j}(x,y;m)]^{(\nu)} =
    \sum_{l=0}^{\nu} {\nu \brack l} q^{(\nu-l)(j-l)} b_j^{(l)}(x,y)*
    a_{n-j}^{(\nu-l)}(x,y;m).
$$
By Lemma~\ref{lemma:special_q-d}, $b_j^{(l)}(x,y;m) = \beta(j,l)
(x-y)^{[j-l]}$ and $a_{n-j}^{(\nu-l)}(x,y;m) =
\beta(n-j,\nu-l)a_{n-j-\nu+l}(x,y;m)$. It can be verified that for
any homogeneous polynomial $b(x,y;m)$ and for any $s \geq 0$,
$(b*a_s)(1,1;m) = q^{ms} b(1,1;m)$. Also, for $x=y=1$,
$b_j^{(l)}(1,1;m) = \beta(j,j) \delta_{j,l}$. We hence have
$\psi_j(1,1) = 0$ for $j>\nu$, and $\psi_j(1,1) = {\nu \brack j}
\beta(j,j) \beta(n-j,\nu-j)q^{m(n-\nu)}$ for $j\leq \nu$. Since
$\beta(n-j,\nu-j) = {n-j \brack \nu-j} \beta(\nu-j,\nu-j)$ and
$\beta(\nu,\nu) = {\nu \brack j} \beta(j,j) \beta(\nu-j,\nu-j)$,
$\psi_j(1,1)= {n-j \brack \nu-j} \beta(\nu,\nu)
q^{m\left(n-\nu\right)}$. Applying $x=y=1$ to the LHS and
rearranging both sides using $\beta(n-i,\nu) = {n-i \brack \nu}
\beta(\nu,\nu)$, we obtain~(\ref{eq:bm_x}).
\end{proof}

Proposition~\ref{prop:bm_x} can be simplified if $\nu$ is less than
the minimum distance of the dual code.

\begin{corollary}\label{cor:binomial_moment_x}
Let $\dr'$ be the minimum rank distance of $\mathcal{C}^{\perp}$. If
$0 \leq \nu < \dr'$, then
\begin{equation}
    \sum_{i=0}^{n-\nu} {n-i \brack \nu} A_i = q^{m(k-\nu)}{n \brack
    \nu}.
\end{equation}
\end{corollary}
\begin{proof}
We have $B_0 = 1$ and $B_1 = \ldots = B_{\nu} = 0$.
\end{proof}

Using the $q^{-1}$-derivative, we obtain another identity.

\begin{proposition}\label{prop:bm_y}
For $0 \leq \nu \leq n$,
\begin{equation}\label{eq:bm_y}
    \sum_{i=\nu}^n {i \brack \nu} q^{\nu(n-i)} A_i = q^{m(k-\nu)}
    \sum_{j=0}^\nu {n-j \brack n-\nu} (-1)^j q^{\sigma_j}
    \alpha(m-j,\nu-j)q^{j(\nu-j)} B_j.
\end{equation}
\end{proposition}

The proof of Proposition~\ref{prop:bm_y} is similar to that of
Proposition~\ref{prop:bm_x}, and is given in
Appendix~\ref{app:prop:bm_y}. Following \cite{macwilliams_77}, we
refer to the LHS of Eqs.~(\ref{eq:bm_x}) and~(\ref{eq:bm_y}) as
binomial moments of the rank distribution of $\mathcal{C}$.
Similarly, when either $\nu$ is less than the minimum distance
$\dr'$ of the dual code, or $\nu$ is greater than the diameter
(maximum distance between any two codewords) $\delta_{\mbox{\tiny
R}}'$ of the dual code, Proposition~\ref{prop:bm_y} can be
simplified.

\begin{corollary}\label{cor:binomial_moment_y}
If $0 \leq \nu < \dr'$, then
\begin{equation}\label{eq:cor_bm_y_d}
    \sum_{i=\nu}^n {i \brack \nu} q^{\nu(n-i)} A_i = q^{m(k-\nu)} {n
    \brack \nu} \alpha(m,\nu).
\end{equation}

For $\delta_{\mbox{\tiny R}}' < \nu \leq n$,
\begin{equation}\label{eq:cor_bm_y_delta}
    \sum_{i=0}^\nu {n-i \brack n-\nu} (-1)^i q^{\sigma_i}
    \alpha(m-i,\nu-i) q^{i(\nu-i)} A_i = 0.
\end{equation}
\end{corollary}

\begin{proof}
Apply Proposition~\ref{prop:bm_y} to $\mathcal{C}$, and use $B_1 =
\ldots = B_\nu = 0$ to prove~(\ref{eq:cor_bm_y_d}). Apply
Proposition~\ref{prop:bm_y} to $\mathcal{C}^\perp$, and use $B_\nu =
\ldots = B_n = 0$ to prove~(\ref{eq:cor_bm_y_delta}).
\end{proof}

\subsection{Pless identities for the rank distribution}
In this section, we consider the analogues of the Pless identities
\cite{macwilliams_77, pless_63}, in terms of Stirling numbers. The
$q$-Stirling numbers of the second kind $S_q(\nu,l)$ are defined
\cite{carlitz_48} to be
\begin{equation}
    \label{eq:S(nu,l)1}
    S_q(\nu,l) \df \frac{q^{-\sigma_l}}{\beta(l,l)} \sum_{i=0}^l
    (-1)^i q^{\sigma_i} {l \brack i} {l-i \brack 1}^\nu,
\end{equation}
and they satisfy
\begin{equation}
    \label{eq:S(nu,l)2}
    {m \brack 1}^\nu = \sum_{l=0}^\nu q^{\sigma_l} S_q(\nu,l) \beta(m,l).
\end{equation}

The following proposition can be viewed as a $q$-analogue of the
Pless identity with respect to $x$ \cite[P$_2$]{pless_63}.

\begin{proposition}\label{prop:pless_x}
For $0 \leq \nu \leq n$,
\begin{equation}\label{eq:pless_x}
    q^{-mk} \sum_{i=0}^n {n-i \brack 1}^\nu A_i = \sum_{j=0}^\nu B_j \sum_{l=0}^\nu {n-j \brack n-l} \beta(l,l)
    S_q(\nu,l) q^{-ml + \sigma_l}.
\end{equation}
\end{proposition}

\begin{proof}
We have
\begin{eqnarray}
    \label{eq:n-i2}
    \sum_{i=0}^n {n-i \brack 1}^\nu A_i &=& \sum_{i=0}^n A_i
    \sum_{l=0}^\nu q^{\sigma_l} S_q(\nu,l) {n-i \brack l} \beta(l,l)\\
    \nonumber
    &=& \sum_{l=0}^\nu q^{\sigma_l} \beta(l,l) S_q(\nu,l) \sum_{i=0}^n {n-i \brack
    l} A_i\\
    \label{eq:n-i1}
    &=& \sum_{l=0}^\nu q^{\sigma_l} \beta(l,l) S_q(\nu,l) q^{m(k-l)} \sum_{j=0}^l {n-j \brack
    n-l} B_j\\
    \nonumber
    &=& q^{mk} \sum_{j=0}^\nu B_j \sum_{l=0}^\nu {n-j \brack n-l} q^{\sigma_l}  \beta(l,l)
    S_q(\nu,l) q^{-ml},
\end{eqnarray}
where~(\ref{eq:n-i2}) follows~(\ref{eq:S(nu,l)2})
and~(\ref{eq:n-i1}) is due to Proposition~\ref{prop:bm_x}.
\end{proof}

Proposition~\ref{prop:pless_x} can be simplified when $\nu$ is less
than the minimum distance of the dual code.

\begin{corollary}\label{cor:pless_x}
For $0 \leq \nu < \dr'$,
\begin{eqnarray}
    \label{eq:cor_pless_x1}
    q^{-mk} \sum_{i=0}^n {n-i \brack 1}^\nu A_i &=&
    \sum_{l=0}^\nu \beta(n,l) S_q(\nu,l) q^{-ml+\sigma_l}\\
    \label{eq:cor_pless_x2}
    &=& q^{-mn} \sum_{i=0}^n {n-i \brack
    1}^\nu {n \brack i} \alpha(m,i).
\end{eqnarray}
\end{corollary}

\begin{proof}
Since $B_0 = 1$ and $B_1 = \cdots = B_\nu = 0$, (\ref{eq:pless_x})
directly leads to~(\ref{eq:cor_pless_x1}). Since the right hand side
of~(\ref{eq:cor_pless_x1}) is transparent to the code, without loss
of generality we choose $\mathcal{C} = \mathrm{GF}(q^m)^n$ and
(\ref{eq:cor_pless_x2}) follows naturally.
\end{proof}

Unfortunately, a $q$-analogue of the Pless identity with respect to
$y$ \cite[P$_1$]{pless_63} cannot be obtained due to the presence of
the $q^{\nu(n-i)}$ term in the LHS of~(\ref{eq:bm_y}). Instead, we
derive its $q^{-1}$-analogue. We denote $p \df q^{-1}$ and define
the functions $\alpha_p(m,u)$, ${n \brack u}_p$, $\beta_p(m,u)$
similarly to the functions introduced in
Section~\ref{sec:notations}, only replacing $q$ by $p$. It is easy
to relate these $q^{-1}$-functions to their counterparts: $
\alpha(m,u) = p^{-mu - \sigma_u}(-1)^u \alpha_p(m,u)$, ${n \brack u}
= p^{-u(n-u)} {n \brack u}_p$, and $\beta(m,u) = p^{-u(m-u) -
\sigma_u} \beta_p(m,u)$.

\begin{proposition}\label{prop:pless_y}
For $0 \leq \nu \leq n$,
\begin{equation}\label{eq:pless_y}
    p^{mk} \sum_{i=0}^n {i \brack 1}_p^\nu A_i = \sum_{j=0}^\nu B_j
    p^{j(m+n-j)} \sum_{l=j}^\nu \beta_p(l,l) S_p(\nu,l) (-1)^l
    {n-j \brack n-l}_p \alpha_p(m-j,l-j).
\end{equation}
\end{proposition}

The proof of Proposition~\ref{prop:pless_y} is given in
Appendix~\ref{app:prop:pless_y}.

\begin{corollary}\label{cor:pless_y}
For $0 \leq \nu < \dr'$,
\begin{equation}\label{eq:pless_y_d}
    p^{mk} \sum_{i=0}^n {i \brack 1}_p^\nu A_i = \sum_{l=0}^\nu \beta_p(n,l) S_p(\nu,l)
    \alpha_p(m,l) (-1)^l.
\end{equation}
\end{corollary}

\begin{proof}
By $B_1 = \ldots = B_\nu = 0$.
\end{proof}

\subsection{Further results on the rank distribution}
For nonnegative integers $\lambda$, $\mu$, and $\nu$, and a linear
code $\mathcal{C}$ with rank weight distribution $\{A_i\}$ we define
\begin{equation}\label{eq:T_lambda_mu_nu}
    T_{\lambda,\mu,\nu} (\mathcal{C}) \df q^{-mk} \sum_{i=0}^n {i \brack
    \lambda}^\mu q^{\nu(n-i)} A_i,
\end{equation}
whose properties are studied below. We refer to
\begin{equation}\label{eq:T_0_0_nu}
    T_{0,0,\nu}(\mathcal{C}) \df q^{-mk} \sum_{i=0}^n q^{\nu(n-i)} A_i
\end{equation}
as the {\em $\nu$-th $q$-moment} of the rank distribution of
$\mathcal{C}$. We remark that for any code $\mathcal{C}$, the $0$-th
order $q$-moment of its rank distribution is equal to $1$. We first
relate $T_{\lambda,1,\nu}(\mathcal{C})$ and
$T_{1,\mu,\nu}(\mathcal{C})$ to $T_{0,0,\nu}(\mathcal{C})$.

\begin{lemma}\label{lemma:T_mu_T_0}
For nonnegative integers $\lambda$, $\mu$, and $\nu$ we have
\begin{eqnarray}
    \label{eq:T_lambda_T_nu}
    T_{\lambda,1,\nu}(\mathcal{C}) &=& \frac{1}{\alpha(\lambda,\lambda)} \sum_{l=0}^\lambda {\lambda \brack l}
    (-1)^l q^{\sigma_l} q^{n(\lambda-l)} T_{0,0,\nu-\lambda+l}(\mathcal{C})\\
    \label{eq:T_mu_T_nu}
    T_{1,\mu,\nu}(\mathcal{C}) &=& (1-q)^{-\mu} \sum_{a=0}^\mu {\mu \choose a} (-1)^a q^{an}
    T_{0,0,\nu-a}(\mathcal{C}).
\end{eqnarray}
\end{lemma}

The proof of Lemma~\ref{lemma:T_mu_T_0} is given in
Appendix~\ref{app:lemma:T_mu_T_0}. We now consider the case where
$\nu$ is less than the minimum distance of the dual code.

\begin{proposition}\label{prop:T_0_0_nu}
For $0 \leq \nu < \dr'$,
\begin{eqnarray}
    \label{eq:T_0_0_nu_3}
    T_{0,0,\nu}(\mathcal{C}) &=& \sum_{j=0}^\nu {\nu \brack j}\alpha(n,j)
    q^{-mj}\\
    \label{eq:T_0_0_nu_1}
    &=& q^{-mn} \sum_{i=0}^n
    {n \brack i} \alpha(m,i) q^{\nu(n-i)}\\
    \label{eq:T_0_0_nu_2}
    &=& q^{-m\nu} \sum_{l=0}^\nu {\nu \brack l} \alpha(m,l)
    q^{n(\nu-l)}.
\end{eqnarray}
\end{proposition}

The proof of Proposition~\ref{prop:T_0_0_nu} is given in
Appendix~\ref{app:prop:T_0_0_nu}. Proposition~\ref{prop:T_0_0_nu}
hence shows that the $\nu$-th $q$-moment of the rank distribution of
a code is transparent to the code when $\nu < \dr'$. As a corollary,
we show that $T_{\lambda,1,\nu}(\mathcal{C})$ and
$T_{1,\mu,\nu}(\mathcal{C})$ are also transparent to the code when
$0 \leq \lambda, \mu \leq \nu < \dr'$.

\begin{corollary}\label{cor:T_1_mu_nu}
For $0 \leq \lambda, \mu \leq \nu < \dr'$,
\begin{eqnarray}
    \label{eq:T_lambda_ct}
    T_{\lambda,1,\nu}(\mathcal{C}) &=& q^{-mn} {n \brack \lambda}
    \sum_{i=\lambda}^n {n-\lambda \brack i-\lambda} q^{\nu(n-i)} \alpha(m,i)\\
    \label{eq:T_mu_ct}
    T_{1,\mu,\nu}(\mathcal{C}) &=& q^{-mn} \sum_{i=0}^n {i \brack 1}^\mu
    q^{\nu(n-i)} {n \brack i} \alpha(m,i).
\end{eqnarray}
\end{corollary}

\begin{proof}
By Lemma~\ref{lemma:T_mu_T_0} and Proposition~\ref{prop:T_0_0_nu},
$T_{\lambda,1,\nu}(\mathcal{C})$ and $T_{1,\mu,\nu}(\mathcal{C})$
are transparent to the code. Thus, without loss of generality we
assume $\mathcal{C} = \mathrm{GF}(q^m)^n$ and~(\ref{eq:T_lambda_ct})
and~(\ref{eq:T_mu_ct}) follow.
\end{proof}

\subsection{Rank weight distribution of MRD codes}
\label{sec:distribution_MRD} The rank weight distribution of linear
Class-I MRD codes was given in \cite{delsarte_78,
gabidulin_pit0185}. Based on our results in
Section~\ref{sec:moments_distrib}, we provide an alternative
derivation of the rank distribution of linear Class-I MRD codes,
which can also be used to determine the rank weight distribution of
Class-II MRD codes.

\begin{proposition}[Rank distribution of linear Class-I MRD
codes]\label{prop:distrib_MRD}

Let $\mathcal{C}$ be an $(n,k,\dr)$ linear Class-I MRD code over
$\mathrm{GF}(q^m)$ $(n \leq m)$, and let $W_\mathcal{C}^{\mbox{\tiny
R}}(x,y) = \sum_{i=0}^n A_i y^i x^{n-i}$ be its rank weight
enumerator. We then have $A_0 = 1$ and for $0 \leq i \leq n-\dr$,
\begin{equation}
    A_{\dr+i} = {n \brack \dr+i} \sum_{j=0}^i (-1)^{i-j}q^{\sigma_{i-j}} {\dr+i \brack
    \dr+j} \left(q^{m(j+1)}-1\right).
\end{equation}
\end{proposition}

\begin{proof}
It can be shown that for two sequences of real numbers
$\{a_j\}_{j=0}^l$ and $\{b_i\}_{i=0}^l$ such that $a_j =
\sum_{i=0}^j {l-i \brack l-j} b_i$ for $0 \leq j \leq l$, we have
$b_i = \sum_{j=0}^i (-1)^{i-j}q^{\sigma_{i-j}} {l-j \brack l-i}a_j$
for $0 \leq i \leq l$.

By Corollary~\ref{cor:binomial_moment_x}, we have $\sum_{i=0}^j
{n-\dr-i \brack n-\dr-j} A_{\dr+i} = {n \brack n-\dr-j} \left(
q^{m(j+1)} - 1\right)$ for $0 \leq j \leq n-\dr$. Applying the
result above to $l=n-\dr$, $a_j = {n \brack n-\dr-j} \left(
q^{m(j+1)}-1 \right)$, and $b_i = A_{\dr+i}$, we obtain
$$
    A_{\dr+i} = \sum_{j=0}^i (-1)^{i-j}q^{\sigma_{i-j}}{n \brack \dr+i}
    {\dr+i \brack \dr+j} \left( q^{m(j+1)}-1 \right).
$$
\end{proof}

We remark that the above rank distribution is consistent with that
derived in \cite{delsarte_78, gabidulin_pit0185}. Since Class-II MRD
codes can be constructed by transposing linear Class-I MRD codes and
the transposition operation preserves the rank weight, the weight
distributions Class-II MRD codes can be obtained accordingly.

\appendix
The proofs in this section use some well-known properties of
Gaussian polynomials \cite{andrews}: ${n \brack k} = {n \brack
n-k}$, ${n \brack k} {k \brack l} = {n \brack l} {n-l \brack n-k}$,
and
\begin{eqnarray}
    \label{eq:binomial_pascal}
    {n \brack k} &=& {n-1 \brack k} + q^{n-k} {n-1 \brack k-1}\\
    \label{eq:binomial_pascal_reversed}
    &=& q^k {n-1 \brack k} + {n-1 \brack k-1}\\
    \label{eq:binomial_n-1}
    &=& \frac{q^n-1}{q^{n-k}-1} {n-1 \brack k}\\
    \label{eq:binomial_k-1}
    &=& \frac{q^{n-k+1}-1}{q^k-1} {n \brack k-1}.
\end{eqnarray}

\subsection{Proof of Lemma~\ref{lemma:Leibniz_x}}
\label{app:lemma:leibniz_x}

We consider homogeneous polynomials $f(x,y;m) = \sum_{i=0}^r f_i y^i
x^{r-i}$ and $u(x,y;m) = \sum_{i=0}^r u_i y^i x^{r-i}$ of degree $r$
as well as $g(x,y;m) = \sum_{j=0}^s g_j y^j x^{s-j}$ and $v(x,y;m) =
\sum_{j=0}^s v_j y^j x^{s-j}$ of degree $s$. First, we need a
technical lemma.

\begin{lemma}\label{lemma:u*v}
If $u_r = 0$, then
\begin{equation}\label{eq:ux}
    \frac{1}{x}(u(x,y;m)*v(x,y;m)) = \frac{u(x,y;m)}{x}*v(x,y;m).
\end{equation}

If $v_s = 0$, then
\begin{equation}\label{eq:vx}
    \frac{1}{x}(u(x,y;m)*v(x,y;m)) = u(x,qy;m)*\frac{v(x,y;m)}{x}.
\end{equation}
\end{lemma}

\begin{proof}
Suppose $u_r=0$, then $\frac{u(x,y;m)}{x} = \sum_{i=0}^{r-1} u_i y^i
x^{r-1-i}$. Hence
\begin{eqnarray*}
    \frac{u(x,y;m)}{x}*v(x,y;m) &=& \sum_{k=0}^{r+s-1}
    \left( \sum_{l=0}^k q^{ls} u_l(m) v_{k-l}(m-l) \right)
    y^k x^{r+s-1-k}\\
    &=& \frac{1}{x}(u(x,y;m)*v(x,y;m)).
\end{eqnarray*}

Suppose $v_s=0$, then $\frac{v(x,y;m)}{x} = \sum_{j=0}^{s-1} v_j y^j
x^{s-1-j}$. Hence
\begin{eqnarray*}
    u(x,qy;m)*\frac{v(x,y;m)}{x} &=& \sum_{k=0}^{r+s-1}
    \left( \sum_{l=0}^k q^{l(s-1)} q^l u_l(m) v_{k-l}(m-l) \right)
    y^k x^{r+s-1-k}\\
    &=& \frac{1}{x}(u(x,y;m)*v(x,y;m)).
\end{eqnarray*}
\end{proof}

We now give a proof of Lemma~\ref{lemma:Leibniz_x}.

\begin{proof}
In order to simplify notations, we omit the dependence of the
polynomials $f$ and $g$ on the parameter $m$. The proof goes by
induction on $\nu$. For $\nu=0$, the result is trivial. For $\nu=1$,
we have
\begin{eqnarray}
    \nonumber
    [f(x,y)*g(x,y)]^{(1)} &=& \frac{1}{(q-1)x} \big[f(qx,y)*g(qx,y) - f(qx,y)*g(x,y) \cdots\\
    \nonumber
    &+& f(qx,y)*g(x,y) - f(x,y)*g(x,y) \big]\\
    \nonumber
    &=& \frac{1}{(q-1)x}\left[ f(qx,y)*(g(qx,y)-g(x,y)) +
    (f(qx,y)-f(x,y))*g(x,y)\right]\\
    \label{eq:x_b}
    &=& f(qx,qy)*\frac{g(qx,y)-g(x,y)}{(q-1)x} +
    \frac{f(qx,y)-f(x,y)}{(q-1)x}*g(x,y)\\
    \label{eq:x_b1}
    &=& q^r f(x,y)*g^{(1)}(x,y) + f^{(1)}(x,y)*g(x,y),
\end{eqnarray}
where~(\ref{eq:x_b}) follows Lemma~\ref{lemma:u*v}.

Now suppose~(\ref{eq:leibniz_nu}) is true for $\nu = {\bar \nu}$. In
order to further simplify notations, we omit the dependence of the
various polynomials in $x$ and $y$. We have
\begin{eqnarray}
    \nonumber
    (f*g)^{({\bar \nu}+1)} &=& \sum_{l=0}^{\bar \nu} {{\bar \nu} \brack l}
    q^{({\bar \nu}-l)(r-l)} \left[ f^{(l)}*g^{({\bar \nu}-l)}\right]^{(1)}\\
    \label{eq:x_c}
    &=& \sum_{l=0}^{{\bar \nu}} {{\bar \nu} \brack l}
    q^{({\bar \nu}-l)(r-l)} \left( q^{r-l}f^{(l)}*g^{({\bar \nu}-l+1)} +
    f^{(l+1)}*g^{({\bar \nu}-l)} \right)\\
    \nonumber
    &=& \sum_{l=0}^{{\bar \nu}} {{\bar \nu} \brack l}
    q^{({\bar \nu}+1-l)(r-l)} f^{(l)}*g^{({\bar \nu}-l+1)} + \sum_{l=1}^{{\bar \nu}+1} {{\bar \nu} \brack l-1}
    q^{({\bar \nu}+1-l)(r-l+1)} f^{(l)}*g^{({\bar \nu}-l+1)}\\
    \nonumber
    &=& \sum_{l=1}^{{\bar \nu}} \left( {{\bar \nu} \brack l} + q^{{\bar \nu}+1-l} {{\bar \nu}
    \brack l-1} \right) q^{({\bar \nu}+1-l)(r-l)} f^{(l)}*g^{({\bar \nu}-l+1)}
    + q^{({\bar \nu}+1)r}f*g^{({\bar \nu}+1)} + f^{({\bar \nu}+1)}*g\\
    \label{eq:x_d}
    &=& \sum_{l=0}^{{\bar \nu}+1} {{\bar \nu}+1 \brack l} q^{({\bar \nu}+1-l)(r-l)}
    f^{(l)}*g^{({\bar \nu}-l+1)},
\end{eqnarray}
where~(\ref{eq:x_c}) follows~(\ref{eq:x_b1}), and~(\ref{eq:x_d})
follows~(\ref{eq:binomial_pascal}).
\end{proof}

\subsection{Proof of Lemma~\ref{lemma:Leibniz_y}}
\label{app:lemma:leibniz_y}

We consider homogeneous polynomials $f(x,y;m) = \sum_{i=0}^r f_i y^i
x^{r-i}$ and $u(x,y;m) = \sum_{i=0}^r u_i y^i x^{r-i}$ of degree $r$
as well as $g(x,y;m) = \sum_{j=0}^s g_j y^j x^{s-j}$ and $v(x,y;m) =
\sum_{j=0}^s v_j y^j x^{s-j}$ of degree $s$. First, we need a
technical lemma.

\begin{lemma}\label{lemma:u*v_y}
If $u_0 = 0$, then
\begin{equation}\label{eq:uy}
    \frac{1}{y}(u(x,y;m))*v(x,y;m)) = q^s \frac{u(x,y;m)}{y}*v(x,y;m-1).
\end{equation}

If $v_0 = 0$, then
\begin{equation}\label{eq:vy}
    \frac{1}{y}(u(x,y;m)*v(x,y;m)) = u(x,qy;m)*\frac{v(x,y;m)}{y}.
\end{equation}
\end{lemma}

\begin{proof}
Suppose $u_0=0$, then $\frac{u(x,y;m)}{y} = \sum_{i=0}^{r-1} u_{i+1}
x^{r-1-i} y^i$. Hence
\begin{eqnarray*}
    q^s \frac{u(x,y;m)}{y}*v(x,y;m-1) &=& q^s \sum_{k=0}^{r+s-1}
    \left( \sum_{l=0}^k q^{ls} u_{l+1} v_{k-l}(m-1-l) \right)
    x^{r+s-1-k}y^k\\
    &=& q^s \sum_{k=1}^{r+s}
    \left( \sum_{l=1}^k q^{(l-1)s} u_l v_{k-l}(m-l) \right)
    x^{r+s-k}y^{k-1}\\
    &=& \frac{1}{y}(u(x,y;m)*v(x,y;m)).
\end{eqnarray*}

Suppose $v_0=0$, then $\frac{v(x,y;m)}{y} = \sum_{j=0}^{s-1} v_{j+1}
x^{s-1-j} y^j$. Hence
\begin{eqnarray*}
    u(x,qy;m)*\frac{v(x,y;m)}{y} &=& \sum_{k=0}^{r+s-1}
    \left( \sum_{l=0}^k q^{l(s-1)} q^l u_l v_{k-l+1}(m-l) \right)
    x^{r+s-1-k}y^k\\
    &=& \sum_{k=1}^{r+s}
    \left( \sum_{l=0}^{k-1} q^{ls} u_l v_{k-l}(m-l) \right)
    x^{r+s-k}y^{k-1}\\
    &=& \frac{1}{y}(u(x,y;m)*v(x,y;m)).
\end{eqnarray*}
\end{proof}

We now give a proof of Lemma~\ref{lemma:Leibniz_y}.

\begin{proof}
The proof goes by induction on $\nu$, and is similar to that of
Lemma~\ref{lemma:Leibniz_x}. For $\nu=0$, the result is trivial. For
$\nu=1$ we can easily show, by using Lemma~\ref{lemma:u*v_y}, that
\begin{equation}
    [f(x,y;m)*g(x,y;m)]^{\{ 1 \}} = f(x,y;m)*g^{\{1\}}(x,y;m) + q^s
    f^{\{1\}}(x,y;m)*g(x,y;m-1).
\end{equation}
It is thus easy to verify the claim by induction on $\nu$.
\end{proof}

\subsection{Proof of Proposition~\ref{prop:P=Q}}\label{app:prop:P=Q}
It was shown in \cite{delsarte_76} that the generalized Krawtchouk
polynomials are the only solutions to the recurrence
\begin{equation}\label{eq:recurrence_Pji}
    P_{j+1}(i+1;m+1,n+1) = q^{j+1} P_{j+1}(i+1;m,n) - q^j
    P_j(i;m,n)
\end{equation}
with initial conditions $P_j(0;m,n) = {n \brack j} \alpha(m,j)$.
Clearly, our polynomials satisfy these initial conditions. We hence
show that $P_j(i;m,n)$ satisfy the recurrence
in~(\ref{eq:recurrence_Pji}). We have
\begin{eqnarray}
    \nonumber
    P_{j+1}(i+1;m+1,n+1) &=& \sum_{l=0}^{i+1} {i+1 \brack l} {n-i \brack
    j+1-l} (-1)^l q^{\sigma_l} q^{l(n-i)} \alpha(m+1-l,j+1-l)\\
    \nonumber
    &=& \sum_{l=0}^{i+1} {i+1 \brack l} {m+1-l \brack
    j+1-l} (-1)^l q^{\sigma_l} q^{l(n-i)} \alpha(n-i,j+1-l)\\
    &=& \sum_{l=0}^{i+1} \left\{ q^l{i \brack l} + {i \brack l-1} \right\}
    \left\{ q^{j+1-l} {m-l \brack j+1-l} + {m-l \brack j-l} \right\}
    \cdots \nonumber \\
    &\cdots& (-1)^l q^{\sigma_l} q^{l(n-i)} \alpha(n-i,j+1-l) \label{eq:P=Q_a} \\
    \nonumber
    &=& \sum_{l=0}^i {i \brack l}
    q^{j+1} {m-l \brack j+1-l} (-1)^l q^{\sigma_l} q^{l(n-i)} \alpha(n-i,j+1-l)\\
    \nonumber
    &+& \sum_{l=0}^i q^l {i \brack l} {m-l \brack j-l}
    (-1)^l q^{\sigma_l} q^{l(n-i)} \alpha(n-i,j+1-l)\\
    \nonumber
    &+& \sum_{l=1}^{i+1} {i \brack l-1}
    q^{j+1-l} {m-l \brack j+1-l} (-1)^l q^{\sigma_l} q^{l(n-i)} \alpha(n-i,j+1-l)\\
    \label{eq:P=Q1}
    &+& \sum_{l=1}^{i+1} {i \brack l-1} {m-l \brack j-l}
    (-1)^l q^{\sigma_l} q^{l(n-i)} \alpha(n-i,j+1-l),
\end{eqnarray}
where~(\ref{eq:P=Q_a}) follows~(\ref{eq:binomial_pascal_reversed}).
Let us denote the four summations in the right hand side
of~(\ref{eq:P=Q1}) as $A$, $B$, $C$, and $D$ respectively. We have
$A = q^{j+1} P_{j+1}(i;m,n)$, and
\begin{eqnarray}
    \label{eq:P=Q_b}
    B &=& \sum_{l=0}^i {i \brack l} {m-l \brack j-l}
    (-1)^l q^{\sigma_l} q^{l(n-i)} \alpha(n-i,j-l) (q^{n-i+l}-q^j),\\
    \nonumber
    C &=& \sum_{l=0}^{i} {i \brack l}
    q^{j-l} {m-l-1 \brack j-l} (-1)^{l+1} q^{\sigma_{l+1}} q^{(l+1)(n-i)}
    \alpha(n-i,j-l)\\
    \label{eq:P=Q_c}
    &=& -q^{j+n-i} \sum_{l=0}^{i} {i \brack l}
    {m-l \brack j-l} (-1)^l q^{\sigma_l} q^{l(n-i)}\alpha(n-i,j-l)
    \frac{q^{m-j}-1}{q^{m-l}-1},\\
    \label{eq:P=Q_d}
    D &=& -q^{n-i} \sum_{l=0}^i {i \brack l} {m-l \brack j-l}
    (-1)^l q^{\sigma_l} q^{l(n-i)} \alpha(n-i,j-l) q^l
    \frac{q^{j-l}-1}{q^{m-l}-1},
\end{eqnarray}
where~(\ref{eq:P=Q_c}) follows~(\ref{eq:binomial_n-1})
and~(\ref{eq:P=Q_d}) follows both~(\ref{eq:binomial_n-1})
and~(\ref{eq:binomial_k-1}).
Combining~(\ref{eq:P=Q_b}),~(\ref{eq:P=Q_c}), and~(\ref{eq:P=Q_d}),
we obtain
\begin{eqnarray*}
    B+C+D &=& \sum_{l=0}^i {i \brack l} {m-l \brack j-l}
    (-1)^l q^{\sigma_l} q^{l(n-i)} \alpha(n-i,j-l) \cdots\\
    &\cdots& \left\{ q^{n-i+l}-q^j - q^{n-i}\frac{q^{m}-q^j}{q^{m-l}-1}
    -q^{n-i} \frac{q^{j}-q^l}{q^{m-l}-1}\right\}\\
    &=& -q^j P_j(i;m,n).
\end{eqnarray*}

\subsection{Proof of Proposition~\ref{prop:bm_y}}
\label{app:prop:bm_y}

Before proving Proposition~\ref{prop:bm_y}, we need two technical
lemmas.

\begin{lemma}\label{lemma:delta}
For all $m$, $\nu$, and $l$, we have
\begin{equation}\label{eq:delta}
    \delta(m,\nu,j) \df \sum_{i=0}^j {j \brack i} (-1)^i
    q^{\sigma_i} \alpha(m-i,\nu) = \alpha(\nu,j) \alpha(m-j,\nu-j) q^{j(m-j)}.
\end{equation}
\end{lemma}

\begin{proof}
The proof goes by induction on $j$. The claim trivially holds for
$j=0$. Let us suppose it holds for $j = {\bar j}$. We have
\begin{eqnarray}
    \nonumber
    \delta(m,\nu,{\bar j}+1) &=& \sum_{i=0}^{{\bar j}+1} {{\bar j}+1 \brack i} (-1)^i
    q^{\sigma_i} \alpha(m-i,\nu)\\
    \label{eq:delta_a}
    &=& \sum_{i=0}^{{\bar j}+1} \left( q^i {{\bar j} \brack i} + {{\bar j} \brack i-1} \right) (-1)^i
    q^{\sigma_i} \alpha(m-i,\nu)\\
    \nonumber
    &=& \sum_{i=0}^{\bar j} q^i {{\bar j} \brack i} (-1)^i q^{\sigma_i}
    \alpha(m-i,\nu) + \sum_{i=1}^{{\bar j}+1} {{\bar j} \brack i-1} (-1)^i
    q^{\sigma_i} \alpha(m-i,\nu)\\
    \nonumber
    &=& \sum_{i=0}^{\bar j} q^i {{\bar j} \brack i} (-1)^i q^{\sigma_i}
    \alpha(m-i,\nu) - \sum_{i=0}^{{\bar j}} {{\bar j} \brack i} (-1)^i
    q^{\sigma_{i+1}} \alpha(m-1-i,\nu)\\
    \nonumber
    &=& \sum_{i=0}^{\bar j} q^i {{\bar j} \brack i} (-1)^i q^{\sigma_i}
    \alpha(m-1-i,\nu-1)q^{m-1-i} (q^\nu - 1)\\
    \nonumber
    &=& q^{m-1}(q^\nu - 1) \delta(m-1,\nu-1,{\bar j})\\
    \nonumber
    &=& \alpha(\nu,{\bar j}+1) \alpha(m-{\bar j}-1,\nu-{\bar j}-1) q^{({\bar j}+1)(m-{\bar
    j}-1)},
\end{eqnarray}
where~(\ref{eq:delta_a})
follows~(\ref{eq:binomial_pascal_reversed}).
\end{proof}

\begin{lemma}\label{lemma:theta}
For all $n$, $\nu$, and $j$, we have
\begin{equation}\label{eq:theta}
    \theta(n,\nu,j) \df \sum_{l=0}^j {j \brack l}
    {n-j \brack \nu-l} q^{l(n-\nu)} (-1)^l q^{\sigma_l}
    \alpha(\nu-l,j-l) = (-1)^j q^{\sigma_j} {n-j \brack n-\nu}.
\end{equation}
\end{lemma}

\begin{proof}
The proof goes by induction on $j$. The claim trivially holds for
$j=0$. Let us suppose it holds for $j = {\bar j}$. We have
\begin{eqnarray}
    \nonumber
    \theta(n,\nu,{\bar j}+1) &=& \sum_{l=0}^{{\bar j}+1} {{\bar j}+1 \brack l}
    {n-1-{\bar j} \brack \nu-l} q^{l(n-\nu)} (-1)^l q^{\sigma_l}
    \alpha(\nu-l,{\bar j}+1-l)\\
    \label{eq:theta1}
    &=& \sum_{l=0}^{{\bar j}+1} \left( {{\bar j} \brack l} + q^{{\bar j}+1-l} {{\bar j} \brack l-1}
    \right) {n-1-{\bar j} \brack \nu-l} q^{l(n-\nu)} (-1)^l q^{\sigma_l}
    \alpha(\nu-l,{\bar j}+1-l)\\
    \nonumber
    &=& \sum_{l=0}^{\bar j} {{\bar j} \brack l} {n-1-{\bar j} \brack \nu-l} q^{l(n-\nu)} (-1)^l q^{\sigma_l}
    \alpha(\nu-l,{\bar j}-l) (q^{\nu-l} - q^{{\bar j}-l})\\
    \label{eq:theta2}
    &+& \sum_{l=1}^{{\bar j}+1} q^{{\bar j}-l+1} {{\bar j} \brack l-1} {n-1-{\bar j} \brack \nu-l} q^{l(n-\nu)} (-1)^l q^{\sigma_l}
    \alpha(\nu-l,{\bar j}-l+1),
\end{eqnarray}
where~(\ref{eq:theta1}) follows~(\ref{eq:binomial_pascal}). Let us
denote the first and second summations in the right hand side of
(\ref{eq:theta2}) as $A$ and $B$, respectively. We have
\begin{eqnarray}
    \nonumber
    A &=& (q^\nu - q^{\bar j}) \sum_{l=0}^{\bar j} {{\bar j} \brack l} {n-1-{\bar j} \brack \nu-l} q^{l(n-1-\nu)} (-1)^l q^{\sigma_l}
    \alpha(\nu-l,{\bar j}-l)\\
    \nonumber
    &=& (q^\nu - q^{\bar j}) \theta(n-1,\nu,{\bar j})\\
    \label{eq:theta_A}
    &=& (q^\nu-q^{\bar j}) (-1)^{\bar j} q^{\sigma_{\bar j}} {n-1-{\bar j} \brack n-1-\nu},
\end{eqnarray}
and
\begin{eqnarray}
    \nonumber
    B &=& \sum_{l=0}^{\bar j} q^{{\bar j}-l} {{\bar j} \brack l} {n-1-{\bar j} \brack \nu-1-l} q^{(l+1)(n-\nu)} (-1)^{l+1} q^{\sigma_{l+1}}
    \alpha(\nu-1-l,{\bar j}-l)\\
    \nonumber
    &=& - q^{{\bar j}+n-\nu} \sum_{l=0}^{\bar j} {{\bar j} \brack l} {n-1-{\bar j} \brack \nu-1-l} q^{l(n-\nu)} (-1)^l q^{\sigma_l}
    \alpha(\nu-1-l,{\bar j}-l)\\
    \nonumber
    &=& -q^{{\bar j}+n-\nu} \theta(n-1,\nu-1,{\bar j})\\
    \label{eq:theta_B}
    &=& -q^{{\bar j}+n-\nu} (-1)^{\bar j} q^{\sigma_{\bar j}} {n-1-{\bar j} \brack n-\nu}.
\end{eqnarray}

Combining~(\ref{eq:theta1}), (\ref{eq:theta_A}),
and~(\ref{eq:theta_B}), we obtain
\begin{eqnarray}
    \nonumber
    \theta(n,\nu,{\bar j}+1) &=& (-1)^{\bar j} q^{\sigma_{\bar j}}
    \left\{ (q^\nu-q^{\bar j}){n-1-{\bar j} \brack n-1-\nu} - q^{{\bar j}+n-\nu} {n-1-{\bar j} \brack n-\nu} \right\}\\
    \label{eq:theta3}
    &=& (-1)^{{\bar j}+1} q^{\sigma_{{\bar j}+1}} {n-1-{\bar j} \brack n-\nu}
    \left\{ - (q^{\nu-{\bar j}}-1) \frac{q^{n-\nu}-1}{q^{\nu-{\bar j}} - 1} + q^{n-\nu}
    \right\}\\
    &=& (-1)^{{\bar j}+1} q^{\sigma_{{\bar j}+1}} {n-1-{\bar j} \brack
    n-\nu},
\end{eqnarray}
where~(\ref{eq:theta3}) follows~(\ref{eq:binomial_k-1}).
\end{proof}

We now give a proof of Proposition~\ref{prop:bm_y}.

\begin{proof}
We apply the $q^{-1}$-derivative with respect to $y$
to~(\ref{eq:before_nu}) $\nu$ times, and we apply $x=y=1$. By
Lemma~\ref{lemma:special_q-1_d} the LHS becomes
\begin{equation}
    \sum_{i=\nu}^n q^{\nu(1-i) + \sigma_\nu} \beta(i,\nu) A_i
    = q^{\nu(1-n) + \sigma_\nu} \beta(\nu,\nu) \sum_{i=\nu}^n
    {i \brack \nu} q^{\nu(n-i)} A_i.
\end{equation}
The RHS becomes $q^{m(k-n)} \sum_{j=0}^n B_j \psi_j(1,1)$, where
\begin{eqnarray}
    \nonumber
    \psi_j(x,y) &\df& \left[ b_j(x,y;m) * a_{n-j}(x,y;m)
    \right]^{ \{\nu\}}\\
    \label{eq:psi1}
    &=& \sum_{l=0}^\nu {\nu \brack l} q^{l(n-j-\nu+l)}
    b_j^{\{l\}}(x,y;m) * a_{n-j}^{\{ \nu-l \}}(x,y;m-l)\\
    \nonumber
    &=& \sum_{l=0}^\nu {\nu \brack l} q^{l(n-j-\nu+l)}
    (-1)^l \beta(j,l) \beta(n-j,\nu-l) q^{-\sigma_{\nu-l}}\cdots\\
    \label{eq:psi2}
    &\cdots& b_{j-l}(x,y;m) * \alpha(m-l,\nu-l)
    a_{n-j-\nu+l}(x,y;m-\nu)\\
    \nonumber
    &=& \beta(\nu,\nu) q^{-\sigma_\nu} \sum_{l=0}^\nu {j \brack l}
    {n-j \brack \nu-l} q^{l(n-j)} (-1)^l q^{\sigma_l}\cdots\\
    \nonumber
    &\cdots& b_{j-l}(x,y;m) * \alpha(m-l,\nu-l)
    a_{n-j-\nu+l}(x,y;m-\nu),
\end{eqnarray}
where~(\ref{eq:psi1}) and~(\ref{eq:psi2}) follow
Lemmas~\ref{lemma:Leibniz_y} and~\ref{lemma:special_q-1_d}
respectively.

We have
\begin{eqnarray}
    \nonumber
    && \left[ b_{j-l} * \alpha(m-l,\nu-l) a_{n-j-\nu+l} \right] (1,1;m-\nu) \cdots\\
    \nonumber
    &=& \sum_{u=0}^{n-\nu} \left[ \sum_{i=0}^u q^{i(n-j-\nu+l)}
    {j-l \brack i} (-1)^i q^{\sigma_i} \alpha(m-i-l,\nu-l) {n-j-\nu+l
    \brack u-i} \alpha(m-\nu-i,u-i) \right]\\
    \nonumber
    &=& q^{(m-\nu)(n-\nu-j+l)} \sum_{i=0}^{j-l} {j-l \brack i}
    (-1)^i q^{\sigma_i} \alpha(m-l-i,\nu-l)\\
    \label{eq:psi_a}
    &=& q^{(m-\nu)(n-\nu-j+l)} \alpha(\nu-l,j-l)
    \alpha(m-j,\nu-j)q^{(j-l)(m-j)},
\end{eqnarray}
where~(\ref{eq:psi_a}) follows Lemma~\ref{lemma:delta}. Hence
\begin{eqnarray}
    \nonumber
    \psi_j(1,1) &=& \beta(\nu,\nu) q^{m(n-\nu) + \nu(1-n) + \sigma_\nu}
    \alpha(m-j,\nu-j)q^{j(\nu-j)}\cdots\\
    \nonumber
    &\cdots&\sum_{l=0}^j {j \brack l}
    {n-j \brack \nu-l} q^{l(n-\nu)} (-1)^l q^{\sigma_l}
    \alpha(\nu-l,j-l)\\
    \label{eq:psi_b}
    &=& \beta(\nu,\nu) q^{m(n-\nu) + \nu(1-n) + \sigma_\nu}
    \alpha(m-j,\nu-j)q^{j(\nu-j)} (-1)^j q^{\sigma_j} {n-j \brack
    n-\nu},
\end{eqnarray}
where~(\ref{eq:psi_b}) follows Lemma~\ref{lemma:theta}.
Incorporating this expression for $\psi_j(1,1)$ in the definition of
the RHS and rearranging both sides, we obtain the result.
\end{proof}

\subsection{Proof of
Proposition~\ref{prop:pless_y}}\label{app:prop:pless_y}

\begin{proof}
Eq.~(\ref{eq:bm_y}) can be expressed in terms of the $\alpha_p(m,u)$
and ${n \brack u}_p$ functions as
\begin{eqnarray}
    \label{eq:pless_y1}
    \sum_{i=\nu}^n {i \brack \nu}_p A_i &=& (-1)^\nu p^{-mk-\sigma_\nu}
    \sum_{j=0}^\nu {n-j \brack n-\nu}_p p^{j(m+n-j)} \alpha_p(m-j,\nu-j)
    B_j.
\end{eqnarray}

We obtain
\begin{eqnarray}
    \label{eq:pless_y2}
    p^{mk} \sum_{i=0}^n {i \brack 1}_p^\nu A_i &=& p^{mk} \sum_{l=0}^\nu p^{\sigma_l} \beta_p(l,l)
    S_p(\nu,l) \sum_{i=l}^n {i \brack l}_p A_i\\
    \label{eq:pless_y3}
    &=& \sum_{l=0}^\nu \beta_p(l,l) S_p(\nu,l) (-1)^l
    \sum_{j=0}^l {n-j \brack n-l}_p p^{j(m+n-j)} \alpha_p(m-j,l-j)
    B_j\\
    \nonumber
    &=& \sum_{j=0}^\nu B_j p^{j(m+n-j)} \sum_{l=j}^\nu \beta_p(l,l) S_p(\nu,l) (-1)^l
    {n-j \brack n-l}_p \alpha_p(m-j,l-j),
\end{eqnarray}
where~(\ref{eq:pless_y2}) and~(\ref{eq:pless_y3})
follow~(\ref{eq:S(nu,l)2}) and~(\ref{eq:pless_y1}) respectively.
\end{proof}

\subsection{Proof of Lemma~\ref{lemma:T_mu_T_0}} \label{app:lemma:T_mu_T_0}

\begin{proof}
We first prove~(\ref{eq:T_lambda_T_nu}):
\begin{eqnarray}
    \label{eq:T_lambda_T_0_1}
    q^{-mk} \sum_{i=0}^n {i \brack \lambda} q^{\nu(n-i)} A_i &=& \frac{q^{-mk}}{\alpha(\lambda,\lambda)}
    \sum_{i=0}^n q^{\nu(n-i)} A_i \sum_{l=0}^\lambda {\lambda \brack l}
    (-1)^l q^{\sigma_l} q^{i(\lambda-l)}\\
    \nonumber
    &=& \frac{q^{-mk}}{\alpha(\lambda,\lambda)} \sum_{l=0}^\lambda {\lambda \brack l}
    (-1)^l q^{\sigma_l} q^{n(\lambda-l)} \sum_{i=0}^n q^{(\nu-\lambda+l)(n-i)} A_i\\
    \nonumber
    &=& \frac{1}{\alpha(\lambda,\lambda)} \sum_{l=0}^\lambda {\lambda \brack l}
    (-1)^l q^{\sigma_l} q^{n(\lambda-l)} T_{0,0,\nu-\lambda+l}(\mathcal{C}),
\end{eqnarray}
where~(\ref{eq:T_lambda_T_0_1}) follows $\alpha(i,\lambda) =
\sum_{l=0}^\lambda {\lambda \brack l} (-1)^l q^{\sigma_l}
q^{i(\lambda-l)}$. We now prove~(\ref{eq:T_mu_T_nu}): since
\begin{equation}
    {i \brack 1}^\mu = \left( \frac{1-q^i}{1-q} \right)^\mu
    = \frac{1}{(1-q)^\mu} \sum_{a=0}^\mu {\mu \choose a} (-1)^a
    q^{ia},
\end{equation}
we obtain
\begin{eqnarray*}
    T_{1,\mu,\nu}(\mathcal{C}) &=& \frac{q^{-mk}}{(1-q)^\mu} \sum_{i=0}^n q^{\nu(n-i)} A_i
    \sum_{a=0}^\mu {\mu \choose a} (-1)^a q^{ia}\\
    &=& \frac{q^{-mk}}{(1-q)^\mu} \sum_{a=0}^\mu {\mu \choose a} (-1)^a q^{an}
    \sum_{i=0}^n q^{(\nu - a)(n-i)} A_i\\
    &=& (1-q)^{-\mu} \sum_{a=0}^\mu {\mu \choose a} (-1)^a q^{an} T_{0,0,\nu-a}(\mathcal{C}).
\end{eqnarray*}
\end{proof}

\subsection{Proof of Proposition~\ref{prop:T_0_0_nu}}
\label{app:prop:T_0_0_nu}

\begin{proof}
From \cite[(3.3.6)]{andrews}, we obtain ${n-i \brack \nu} =
\frac{1}{\alpha(\nu,\nu)} \sum_{l=0}^\nu {\nu \brack l} (-1)^{\nu-l}
q^{\sigma_{\nu-l}} q^{l(n-i)}$, and hence
\begin{eqnarray}
    \nonumber
    q^{-mk} \sum_{i=0}^n {n-i \brack \nu} A_i &=& q^{-mk} \sum_{i=0}^n
    A_i \frac{1}{\alpha(\nu,\nu)} \sum_{l=0}^\nu {\nu \brack l}
    (-1)^{\nu-l} q^{\sigma_{\nu-l}} q^{l(n-i)}\\
    \nonumber
    &=& \frac{q^{-mk}}{\alpha(\nu,\nu)}  \sum_{l=0}^\nu {\nu \brack l}
    (-1)^{\nu-l} q^{\sigma_{\nu-l}} \sum_{i=0}^n q^{l(n-i)} A_i\\
    \label{eq:T1}
    &=& \frac{1}{\alpha(\nu,\nu)} \sum_{l=0}^\nu {\nu \brack l}
    (-1)^{\nu-l} q^{\sigma_{\nu-l}} T_{0,0,l}(\mathcal{C}),
\end{eqnarray}
where~(\ref{eq:T1}) follows~(\ref{eq:T_0_0_nu}). By
Corollary~\ref{cor:binomial_moment_x}, we have for $\nu < \dr'$,
$\sum_{l=0}^\nu {\nu \brack l}  (-1)^{\nu-l} q^{\sigma_{\nu-l}}
T_{0,0,l}(\mathcal{C}) = q^{-m\nu} \alpha(n,\nu)$, and we obtain
\begin{eqnarray}
    \nonumber
    \sum_{j=0}^\nu {\nu \brack j}\alpha(n,j) q^{-mj} &=&
    \sum_{j=0}^\nu {\nu \brack j} \sum_{l=0}^j {j \brack l}
    (-1)^{j-l} q^{\sigma_{j-l}} T_{0,0,l}(\mathcal{C})\\
    \nonumber
    &=& \sum_{l=0}^\nu T_{0,0,l}(\mathcal{C}) {\nu \brack l} \sum_{j=0}^\nu
    {\nu-l \brack j-l} (-1)^{j-l} q^{\sigma_{j-l}}\\
    \label{eq:T2}
    &=& T_{0,0,\nu}(\mathcal{C}),
\end{eqnarray}
where~(\ref{eq:T2}) follows $\sum_{j=0}^{\nu-l} {\nu-l \brack j}
(-1)^j q^{\sigma_j} = \delta_{\nu,l}$, which in turn is a special
case of \cite[(3.3.6)]{andrews}. This proves~(\ref{eq:T_0_0_nu_3}).
Thus, $T_{0,0,\nu}(\mathcal{C})$ is transparent to the code, and
(\ref{eq:T_0_0_nu_1}) can be shown by choosing $\mathcal{C} =
\mathrm{GF}(q^m)^n$ without loss of generality.

Suppose $S(\nu,n,m) \df \sum_{j=0}^\nu {\nu \brack j}\alpha(n,j)
q^{-mj}$, then $S(\nu,n,m) = S(n,\nu,m)$ since ${\nu \brack
j}\alpha(n,j) = {n \brack j}\alpha(\nu,j)$. Also,
combining~(\ref{eq:T_0_0_nu_3}) and~(\ref{eq:T_0_0_nu_1}) yields
$S(\nu,n,m) = q^{n(\nu-m)}S(n,m,\nu)$. Therefore, we obtain
$S(\nu,n,m) = q^{\nu(n-m)} S(\nu,m,n)$, which
proves~(\ref{eq:T_0_0_nu_2}).
\end{proof}

\bibliographystyle{IEEETran}
\bibliography{gpt}

\end{document}